\documentclass[aps,prl,twocolumn,10pt]{revtex4-1}
\usepackage{graphicx}
\usepackage{latexsym}
\usepackage{amsmath}
\usepackage{bm}
\newcommand{\beq}{\begin{equation}}
\newcommand{\eeq}{\end{equation}}
\newcommand{\ba}{\begin{array}{ccc}}
\newcommand{\ea}{\end{array}}

\newcommand{\bK}{{\bm K}}
\newcommand{\bQ}{{\bm Q}}
\newcommand{\bra}[1]{\left\langle #1\right|}
\newcommand{\ket}[1]{\left| #1\right\rangle}

\newcommand{\bvec}[1]{\boldsymbol{#1}}
\newcommand{\ii}{\mathrm i\,}
\def\bea{\begin{eqnarray}}
\def\eea{\end{eqnarray}}
\usepackage{color}

\begin{document}
\title{Bond order instabilities in a correlated two-dimensional metal}

\author{Andrea Allais}
\affiliation{Department of Physics, Harvard University, Cambridge MA 02138}

\author{Johannes Bauer}
\affiliation{Department of Physics, Harvard University, Cambridge MA 02138}

\author{Subir Sachdev}
\affiliation{Department of Physics, Harvard University, Cambridge MA
02138}

\date{\today}

\begin{abstract}
Motivated by recent experimental evidence of charge order in the pseudogap
phase of cuprates, we perform a variational analysis of charge-neutral,
spin-singlet ordering in metals on the square lattice, using a wavefunction with double occupancy projected out. 
We examine ordering with and without time-reversal
symmetry, with arbitrary wavevector and tunable form factor. Depending on
parameters, we find $d$-wave bond  density wave ordering with wavevector
either parallel to the lattice generators or diagonally oriented, or a ground
state which carries a time reversal-breaking pattern of spontaneous currents. 
\end{abstract}

\maketitle

\begin{figure}
\includegraphics[width=180pt]{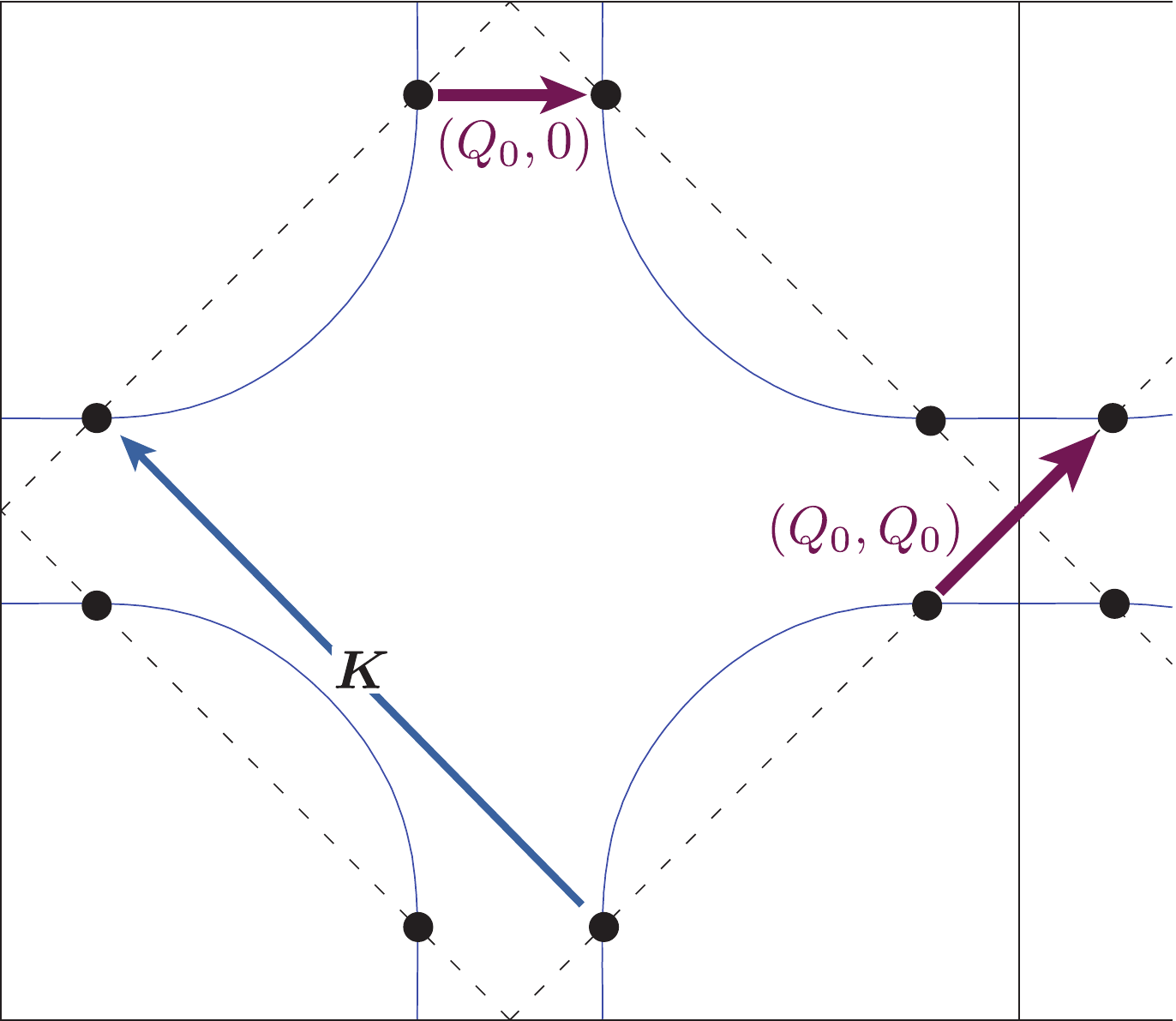}
\caption{Fermi surface with $t_1 = 1$, $t_2 = -0.32$, $t_3=0.128$, and $\mu=-1.11856$.
For this dispersion we have $Q_0 = 4 \pi/11$.
}
\label{fig:bz}
\end{figure}

There is growing experimental evidence that charge density wave order is a
generic feature of underdoped cuprate high temperature superconductors. Its
presence has been long established in $\rm{La}_2 \rm{Cu} \rm{O}_4$ based
compounds \cite{tranquada, abbamonte}. In BSCCO, periodic modulations in the
local density of states have been detected with STM, both in the mixed state
near the vortex cores \cite{hoffman}, and in the pseudogap state
\cite{vershinin}. In $\rm{Y} \rm{Ba}_2 \rm{Cu}_3\rm{O}_\gamma$,  long range,
static charge order has been detected with NMR \cite{julien,julienvortex}, and
its effects show up in  thermodynamic properties \cite{proust}. This order may
explain the quantum oscillations seen in high magnetic field \cite{doiron,sebastian}. At
zero field, incommensurate charge order has been detected with resonant
\cite{keimer, hawthorn} and hard \cite{chang} X-ray scattering. Collectively,
the experiments point to the existence of  
incommensurate charge correlations in the CuO$_2$ plane, which are stabilized
to static, long range order by a magnetic field, and in general compete with
superconductivity. The wavevector is consistently found to be directed along
the copper-oxygen bonds, and  
appears to be related to geometric properties of the Fermi surface
\cite{comin} (Fig~\ref{fig:bz}). There are also indications that the charge 
order lies predominantly on the {\em bonds\/} connecting the Cu sites \cite{ssrmp,kohsaka,hawthorn2}.

Motivated by these remarkable experimental developments, we present a thorough
exploration of charge density wave (CDW) instabilities in a strongly
correlated, antiferromagnetic metal as described by a variational wavefunction
in which doubly-occupied sites are projected out.  
Our main conclusions are that i) the antiferromagnetic interaction and electron correlations
can cause the condensation of a $d$-wave CDW with the experimentally observed
wavevector (we define a $d$-wave CDW as in Refs.~\cite{max,rolando}---see below), 
a result that proved elusive so far under controlled
approximations, and ii) for a different range of parameters, a state is
favored which supports time-reversal breaking permanent currents, known 
in the literature as the staggered flux (SF) state \cite{marston,kotliar,sudip,leewen,laughlin}.

From a theoretical point of view
\cite{max,metzner,yamase,pepin,kee,bulut,rolando,dhlee,jay,pepin2,chubukov},  
a charge/bond density wave is a natural instability of a metal with
antiferromagnetic interactions. This was explored in the context of a weak
coupling analysis in Ref.~\onlinecite{rolando}, which showed that a
bond-density wave arises with a wavevector $\bQ$ and a local $d$-wave pattern
of the bond modulations. Consistent with expectations, the wavevector $\bQ$
was found to be very close to that connecting two hot spots on the Fermi surface (FS), which are points
of the FS connected by the antiferromagnetic wavevector $\bK = (\pi, \pi)$
(Fig.~\ref{fig:bz}). However, the globally optimal wavevector was found to run
diagonally.  
A restricted optimization for wavevectors parallel to the copper-oxygen bonds
\cite{rolando} (the wavevector direction observed in experiments)  
also yielded a wavevector close to that connecting hot spots (Fig.~\ref{fig:bz}) and
with a form factor which remained predominantly $d$-wave.

Here we show how strong electron correlations modify the picture above. We account
for those Mott correlations variationally, using a Gutzwiller projected wavefunction
which suppresses double occupancy of all sites
\cite{gutz,gros,arun,raczkowski,trivedi,galatski}. We allow the condensation of a charge
density waves with arbitrary wavevector and with tunable form factor (the form
factors determines the intra-unit-cell structure of the density wave), as well
as states which break time-reversal symmetry and carry different patterns of
spontaneous currents. Among previously studied states, our study includes
charged stripes \cite{stripes}, bond density waves \cite{ssrmp,vojta4},
Ising-nematic order \cite{YK00,HM00,OKF01}, staggered flux states
\cite{marston,kotliar,sudip,leewen,laughlin}, and states with spontaneous currents \cite{varma}. Our main
results are summarized  
below in Fig.~\ref{fig:variational}: we find regimes where the globally optimal state has a CDW with wavevector very close to 
$(Q_0,0)$ and a $d$-wave form factor.

We consider the following model of a metal with antiferromagnetic and Coulomb interactions,
\begin{equation}
\begin{split}
H =  \sum_{\bvec x,\bvec a}\Big[&- t_{a} c_{\bvec x+\bvec a}^\dagger c^{\phantom{\dagger}}_{\bvec x} +  \frac{J_{a}}{8} c^\dagger_{\bvec x+\bvec a} \vec{\sigma} c^{\phantom{\dagger}}_{\bvec x + \bvec a}\cdot c^\dagger_{\bvec x}\vec{\sigma} c^{\phantom{\dagger}}_{\bvec x}\\
& + \frac{V_{a}}{2} c_{\bvec x+\bvec a}^\dagger c^{\phantom{\dagger}}_{\bvec x+\bvec a} c_{\bvec x}^\dagger c^{\phantom{\dagger}}_{\bvec x}\Big] + H_U \,, \label{ham}
\end{split}
\end{equation}
where the electrons $c$ live on the sites $\bvec x$ of a square lattice with
spacing $a=1$ and $\vec\sigma$ is the vector of Pauli matrices, which act on an implicit spin
index. The vector $\bvec a$ connects neighboring sites, and we allow first,
second, and third neighbor hoppings $t_1$, $t_2$, $t_3$,
respectively. However,  we limit the Coulomb ($V$) and exchange ($J)$
interactions to the nearest neighbors. $H_U$ represents an 
infinite on-site Coulomb repulsion, which we account for by projecting out
doubly occupied sites. The couplings are real and preserve all lattice
symmetries, and we work at fixed density. 

We seek to minimize the energy of the Hamiltonian (\ref{ham}) within the space of states
\begin{equation}\label{varstate}
\ket{\text{var}} \equiv \left[\prod_{i}\left(1 - n_{i\uparrow}n_{i\downarrow}\right)\right]\ket{\text{gd}(H_\text{var})}\,,
\end{equation} 
where $\ket{\text{gd}(H_\text{var})}$ is the ground state of the quadratic hamiltonian
\begin{equation}
H_\text{var} = \sum_{\bvec x, \bvec a} \left[-T_{\bvec a} + \phi_{\bvec a} \cos\bvec Q\cdot(\bvec x + \bvec a/2)\right] c_{\bvec x+\bvec a}^\dagger c^{\phantom{\dagger}}_{\bvec x}\,, \label{Hvar}
\end{equation}
and the state $\ket{\text{var}}$ has no doubly occupied sites by
construction. $T_{\bvec a}$ are variational hopping parameters and
$\phi_{\bvec a}$ parameters for the ordering wave-function.
We note that the particular parameterization in Eq.~(\ref{Hvar}) is carefully
chosen \cite{max,rolando}: it is crucial that $\bvec Q$ couple to the center-of-mass co-ordinate
of the particle-hole pair for an efficient symmetry characterization of order parameters
at incommensurate $\bvec Q$. Previous analyses \cite{chetan,vojta4,sudip} did not make this choice.
 
\begin{figure}
 \begin{center}
  \mbox{%
  a)\includegraphics[width=110pt]{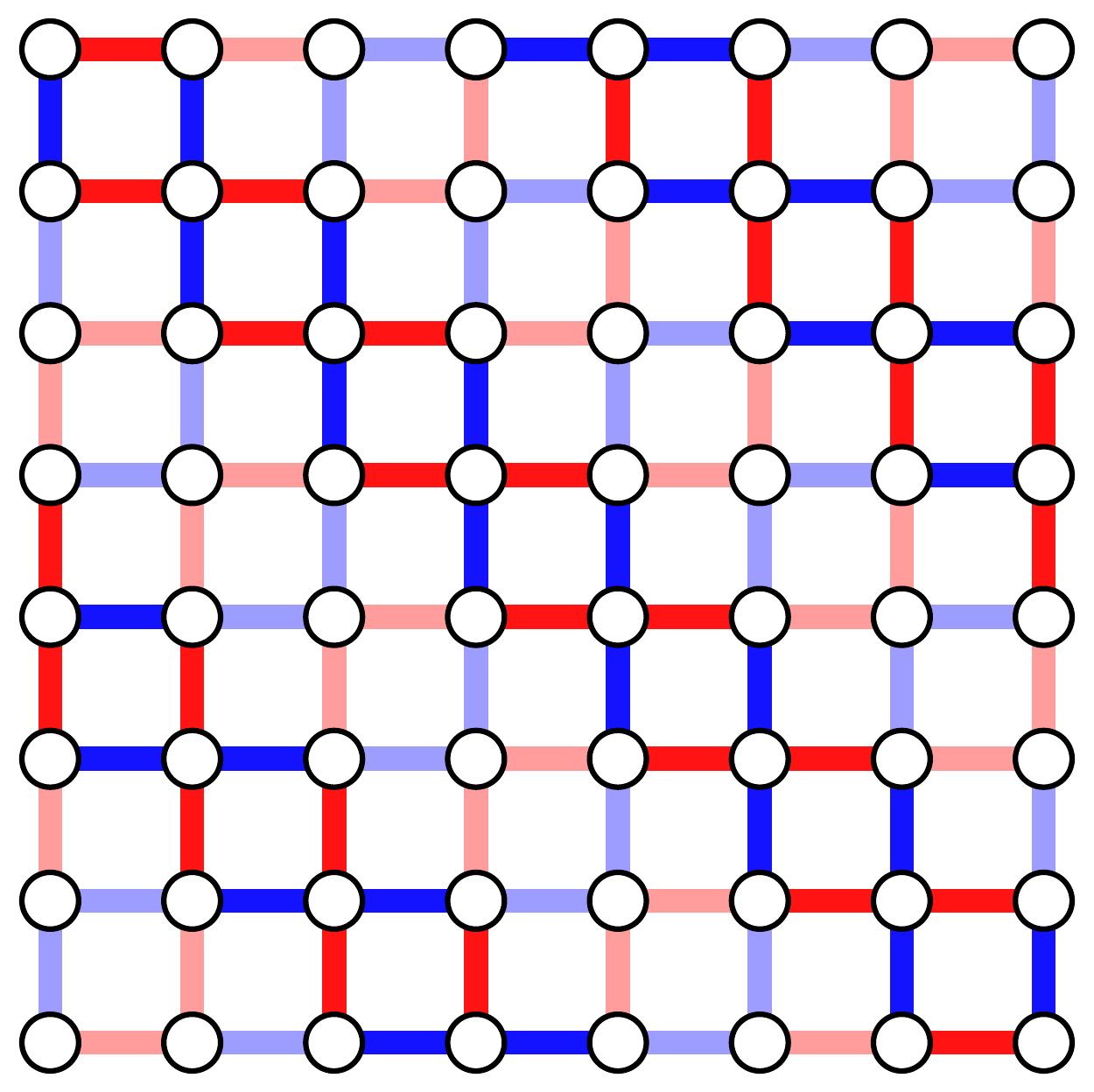}%
  b)\includegraphics[width=110pt]{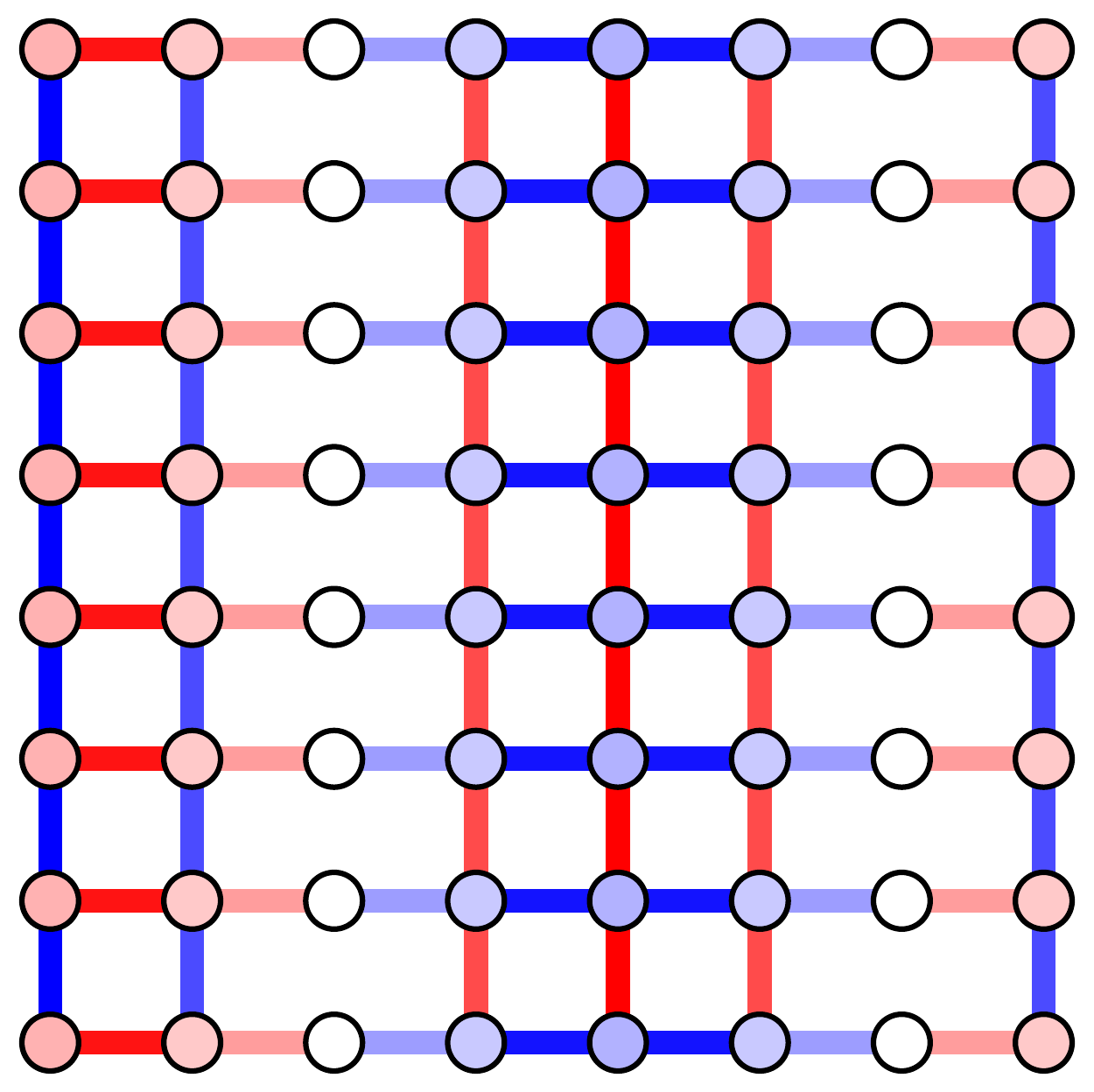}}\\~\\
  \mbox{c)\includegraphics[width=110pt]{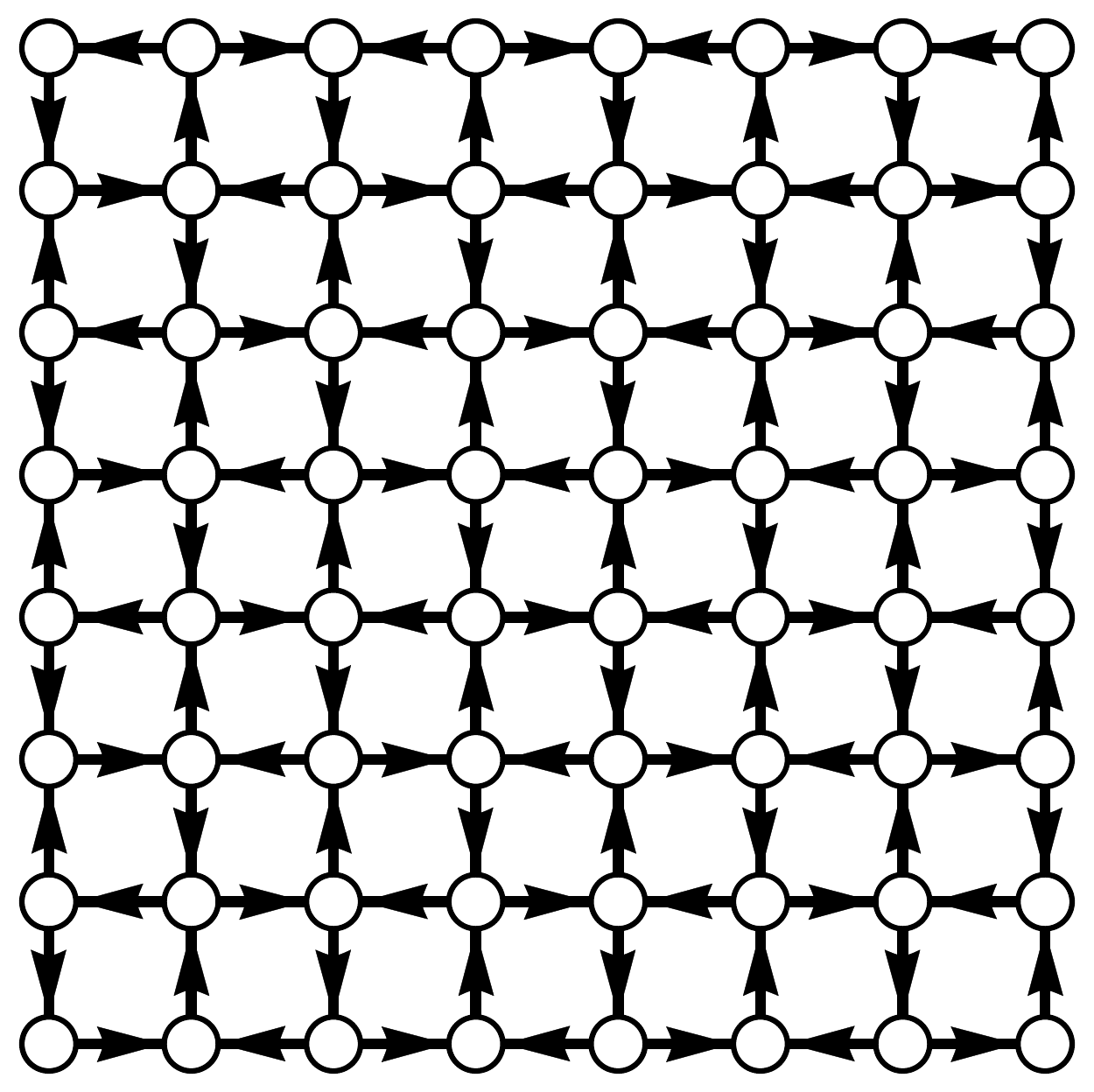}%
}%
 \end{center}
\caption{Schematic illustration of the ordering patterns in real space. a)
  diagonal $d$-wave CDW ($P_{3}\neq 0$, $\bQ = (0.25,0.25)\pi$) b) horizontal,
  predominantly $d$-wave CDW ($P_{3}\neq 0$, $\bQ = (0.25,0)\pi$) c) staggered
  flux state ($P_5\neq0$, $\bQ = (\pi,\pi)$). For the time reversal invariant
  orders a) and b) we show the fluctuation of $\langle c^\dag_{\bvec x+\bvec
    a} c_{\bvec x}\rangle$ about the average with positive (red) and negative
  (blue) values. For the time reversal breaking
  order c) we show the pattern of permanent currents.} 
\label{fig:real_space_orders}
\end{figure}

This variational ansatz allows for the condensation of a charge/bond density
wave with wavevector $\bvec Q$, whose local pattern (the form factor) is
determined by the wavefunction $\phi$. That is, 
\begin{equation}
 \left\langle c^\dag_{\bvec x + \bvec a}c^{\phantom{\dag}}_{\bvec x}\right\rangle_{\text{var}} = -\bar T_{\bvec a} + \bar \phi_{\bvec a} \cos\left[\bvec Q\cdot(\bvec x + \bvec a/2)\right]\,,
\end{equation} 
where $\bar \phi$ is nonzero if and only if $\phi$ is nonzero, and they both
have the same symmetries under point group transformations and time
reversal. We further restrict $\phi_{\bvec a}$ to have the form 
\begin{equation}
\begin{split}
 \phi_{\bvec a} = P_1 \delta_{\bvec a} &+\phantom{\ii} P_2 (\delta_{\bvec a -
   \hat{\bvec{x}}} + \delta_{\bvec a - \hat{\bvec{y}}} + \delta_{\bvec a +
   \hat{\bvec{x}}} +\delta_{\bvec a + \hat{\bvec{y}}})\\ 
 &+\phantom{\ii} P_3 (\delta_{\bvec a - \hat{\bvec{x}}} - \delta_{\bvec a -
   \hat{\bvec{y}}} + \delta_{\bvec a + \hat{\bvec{x}}} -\delta_{\bvec a +
   \hat{\bvec{y}}})\\ 
 &+\ii P_4 (\delta_{\bvec a + \hat{\bvec{x}}} + \delta_{\bvec a -
   \hat{\bvec{y}}} - \delta_{\bvec a + \hat{\bvec{x}}} -\delta_{\bvec a +
   \hat{\bvec{y}}})\\ 
 &+\ii P_5 (\delta_{\bvec a - \hat{\bvec{x}}} - \delta_{\bvec a -
   \hat{\bvec{y}}} - \delta_{\bvec a + \hat{\bvec{x}}} +\delta_{\bvec a +
   \hat{\bvec{y}}})\,,
\end{split}
\end{equation} 
with $\hat{\bvec{x}}=(1,0)$, $\hat{\bvec{y}}=(0,1)$.
Time reversal is preserved if $P_4$ = $P_5$ = 0. If $P_1, P_2 \gg P_3$, a
predominantly $s$-wave charge/bond density wave is induced, whereas $P_1, P_2
\ll P_3$ induces a predominantly $d$-wave bond density wave. It is important
to notice that, for generic $\bvec Q$, $s$- and $d$-wave characters mix, so
that $P_3 = 0$ does not imply that $\bar \phi_{\bvec a}$ is purely
$s$-wave. If $P_4$ or $P_5$ are nonzero, a time reversal breaking pattern of
spontaneous currents is created. In particular the state with only $P_5\neq0$
and $\bQ = (\pi, \pi)$ is the staggered flux state. Figure
\ref{fig:real_space_orders} shows a schematic representation of a few relevant
CDW and spontaneous current patterns. 

\begin{figure*}
\includegraphics[width=140.8pt]{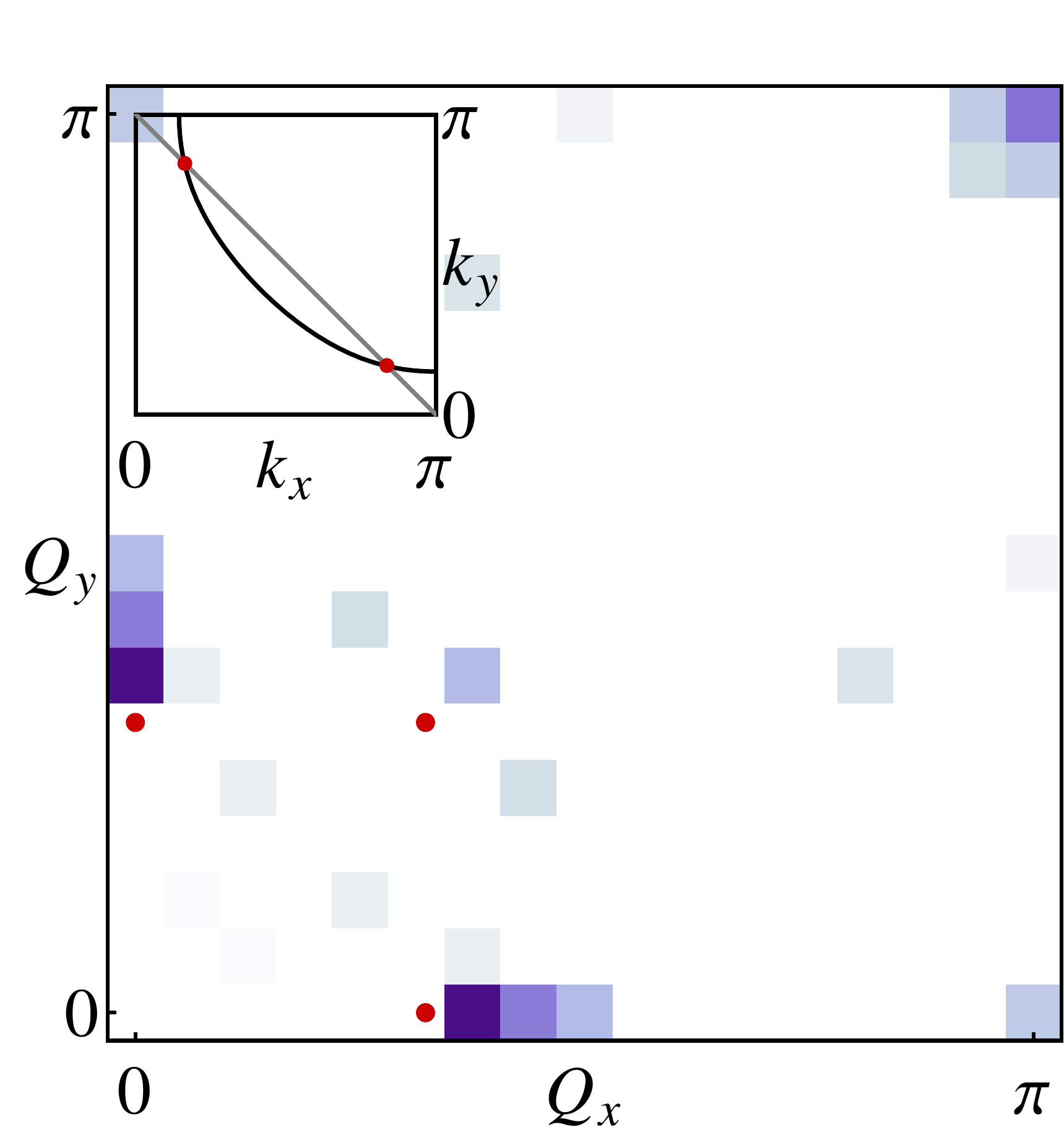}
\includegraphics[width=24.2pt]{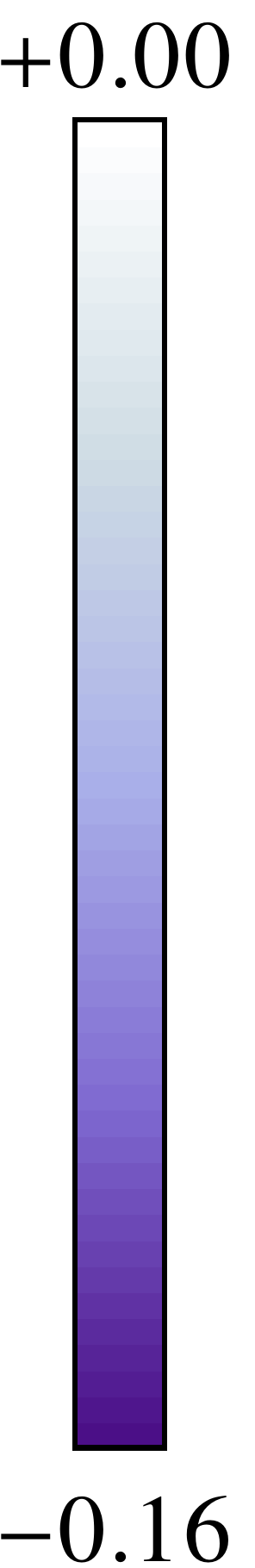}
\includegraphics[width=140.8pt]{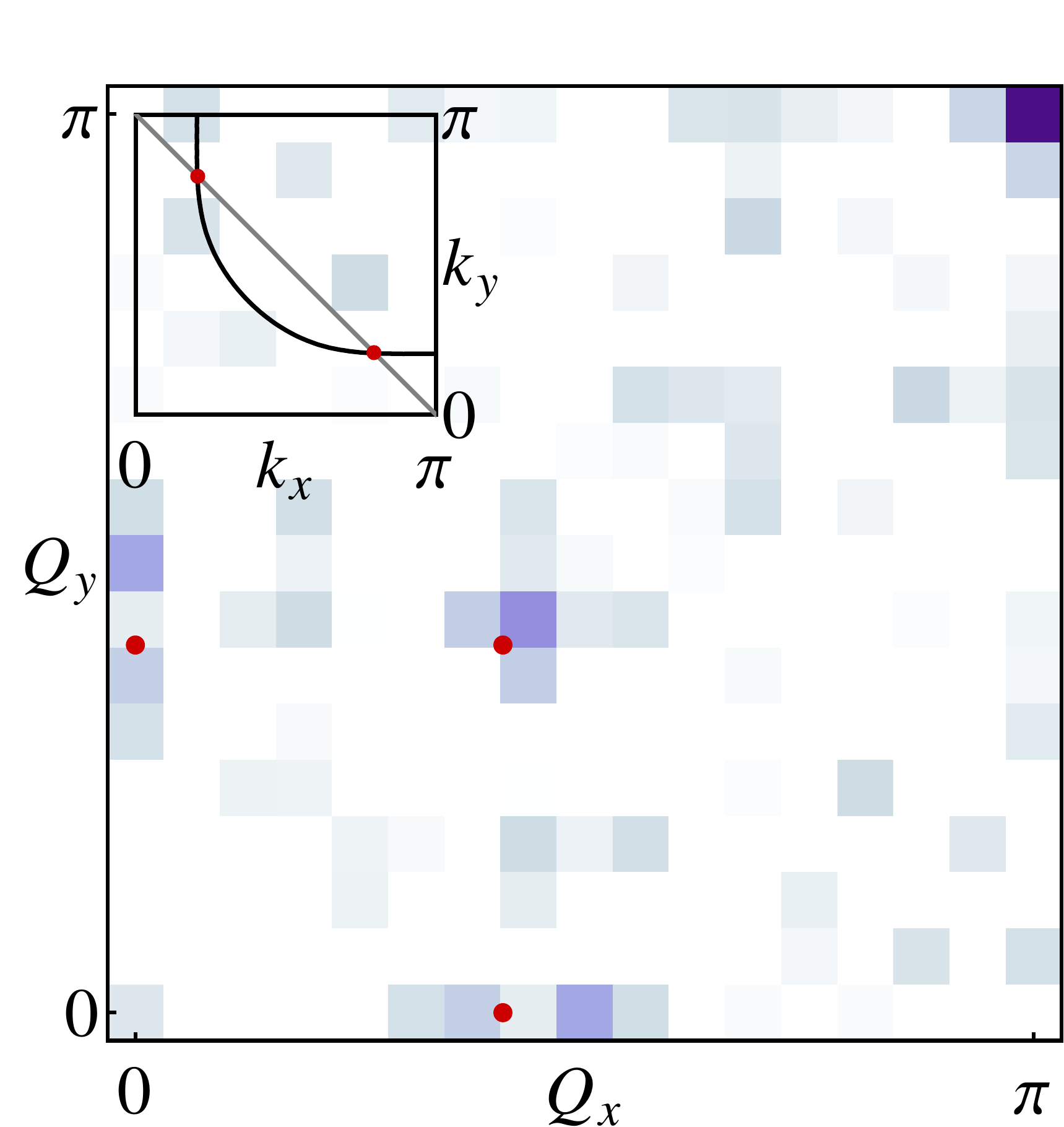}
\includegraphics[width=24.2pt]{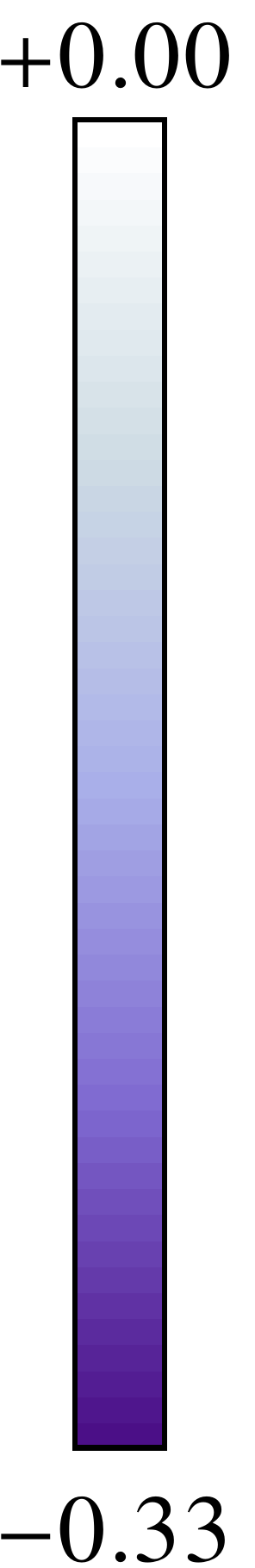}
\includegraphics[width=140.8pt]{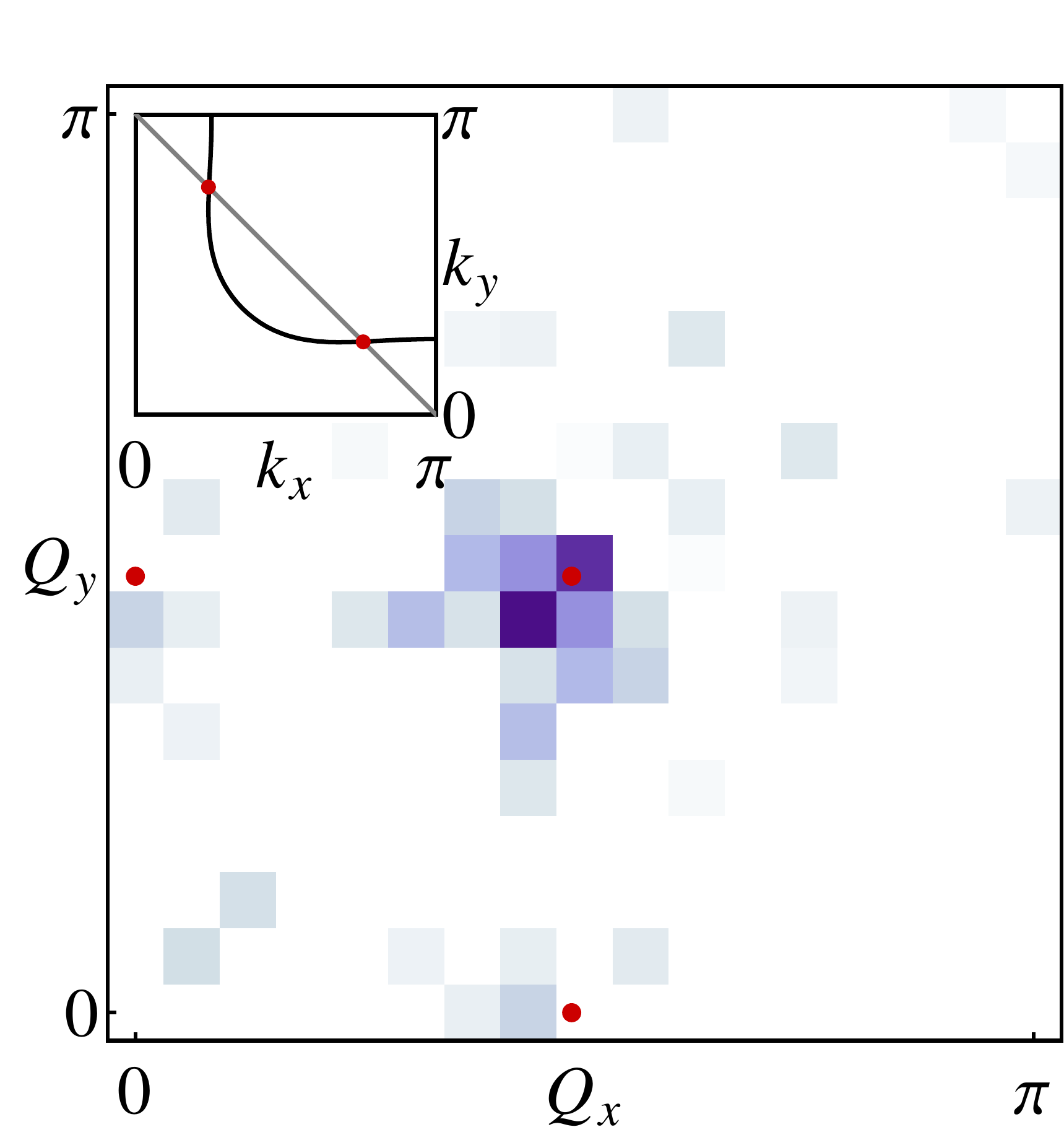}
\includegraphics[width=24.2pt]{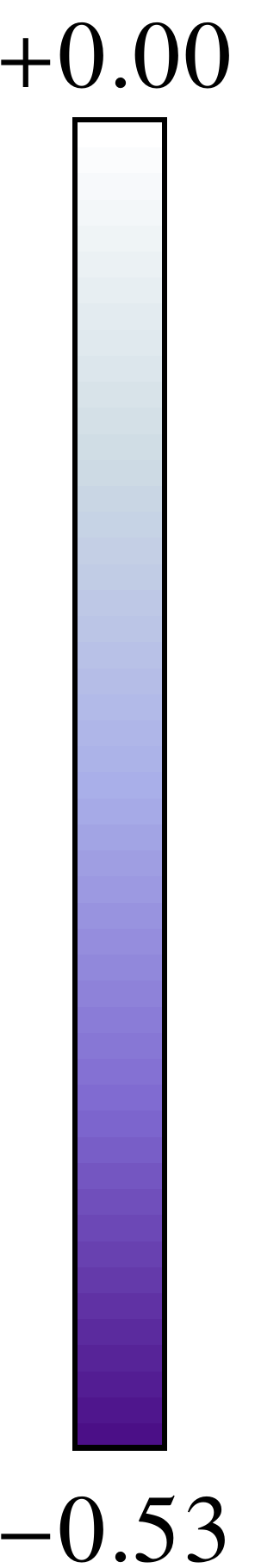}\\~\\
\includegraphics[width=165pt]{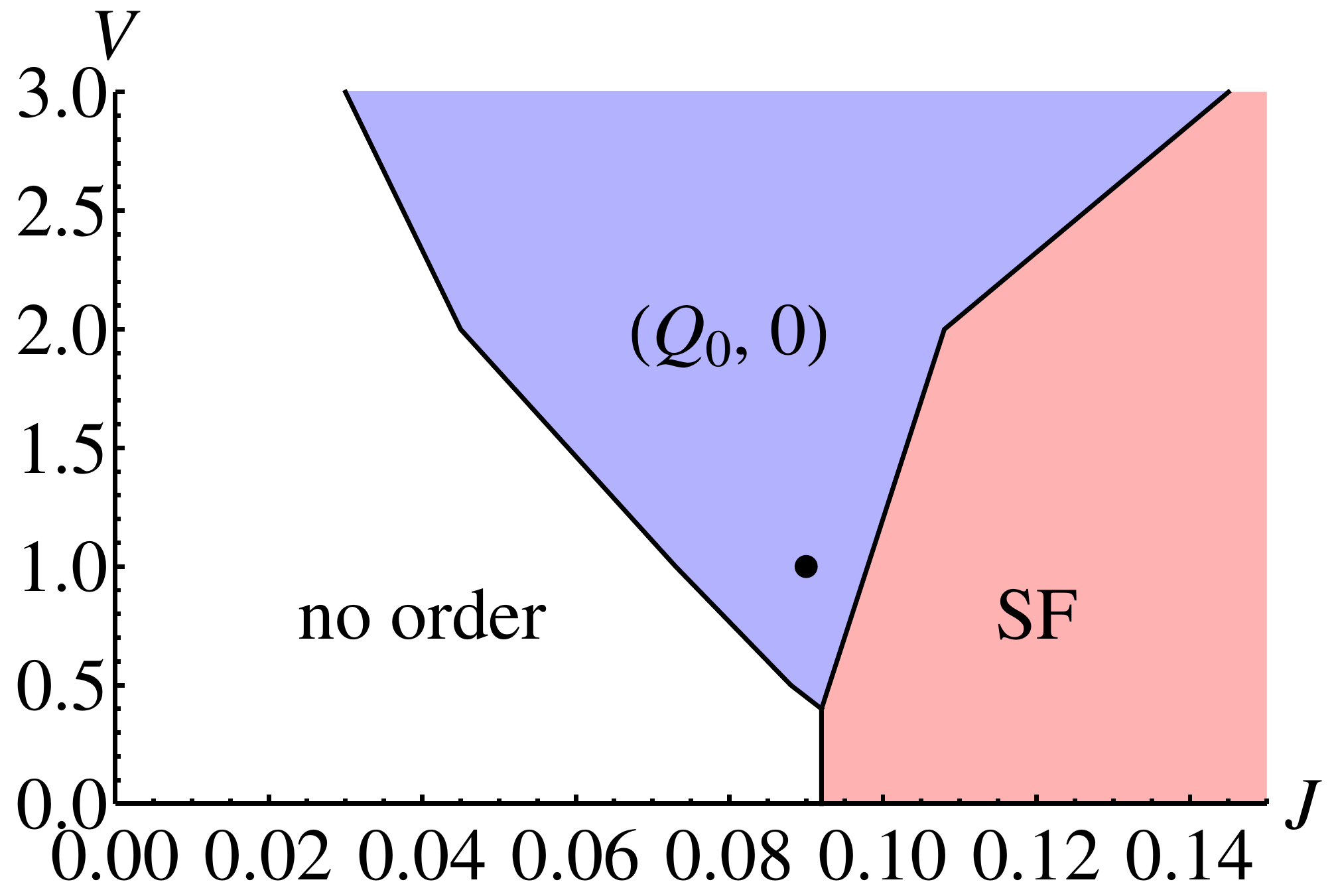}
\includegraphics[width=165pt]{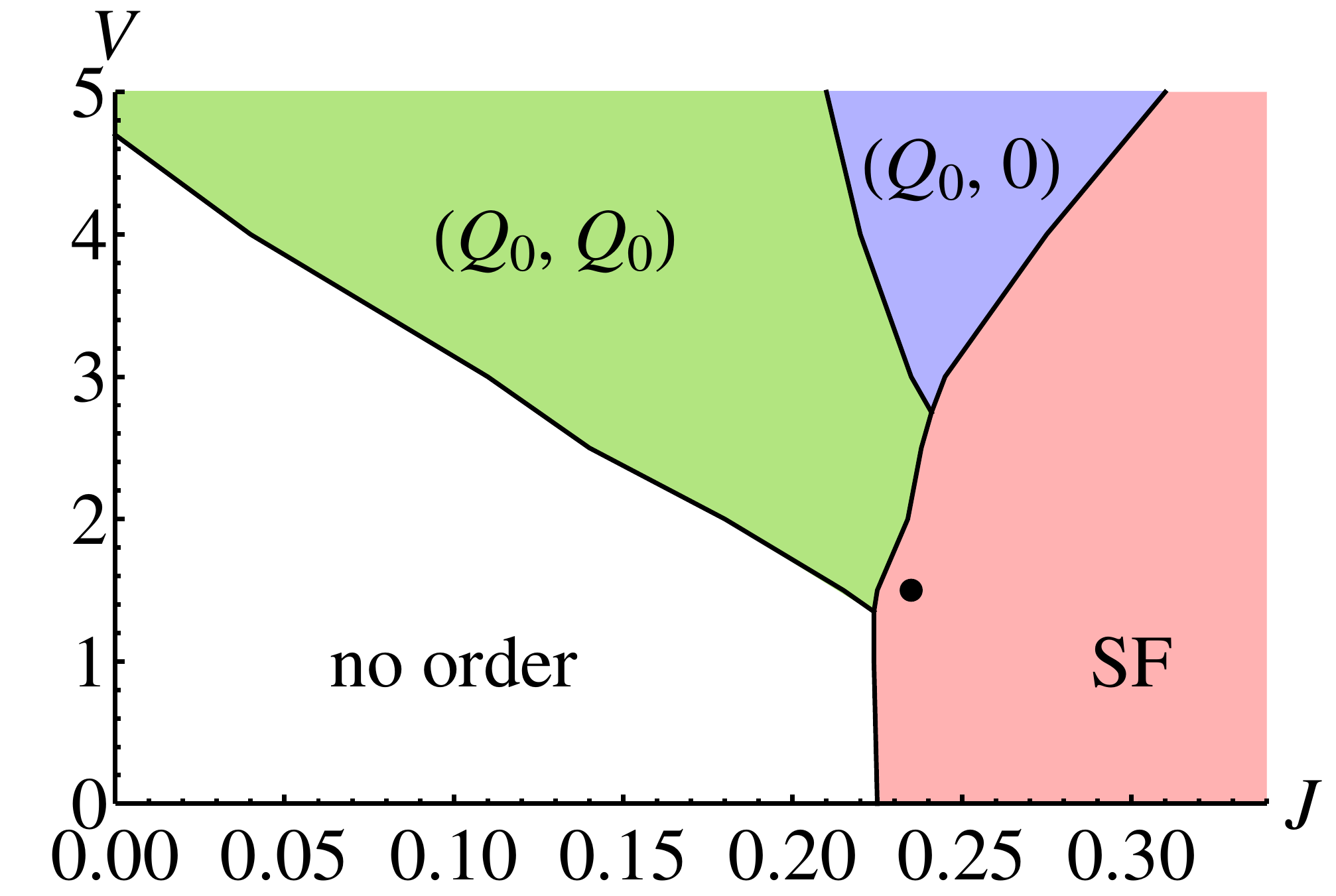}
\includegraphics[width=165pt]{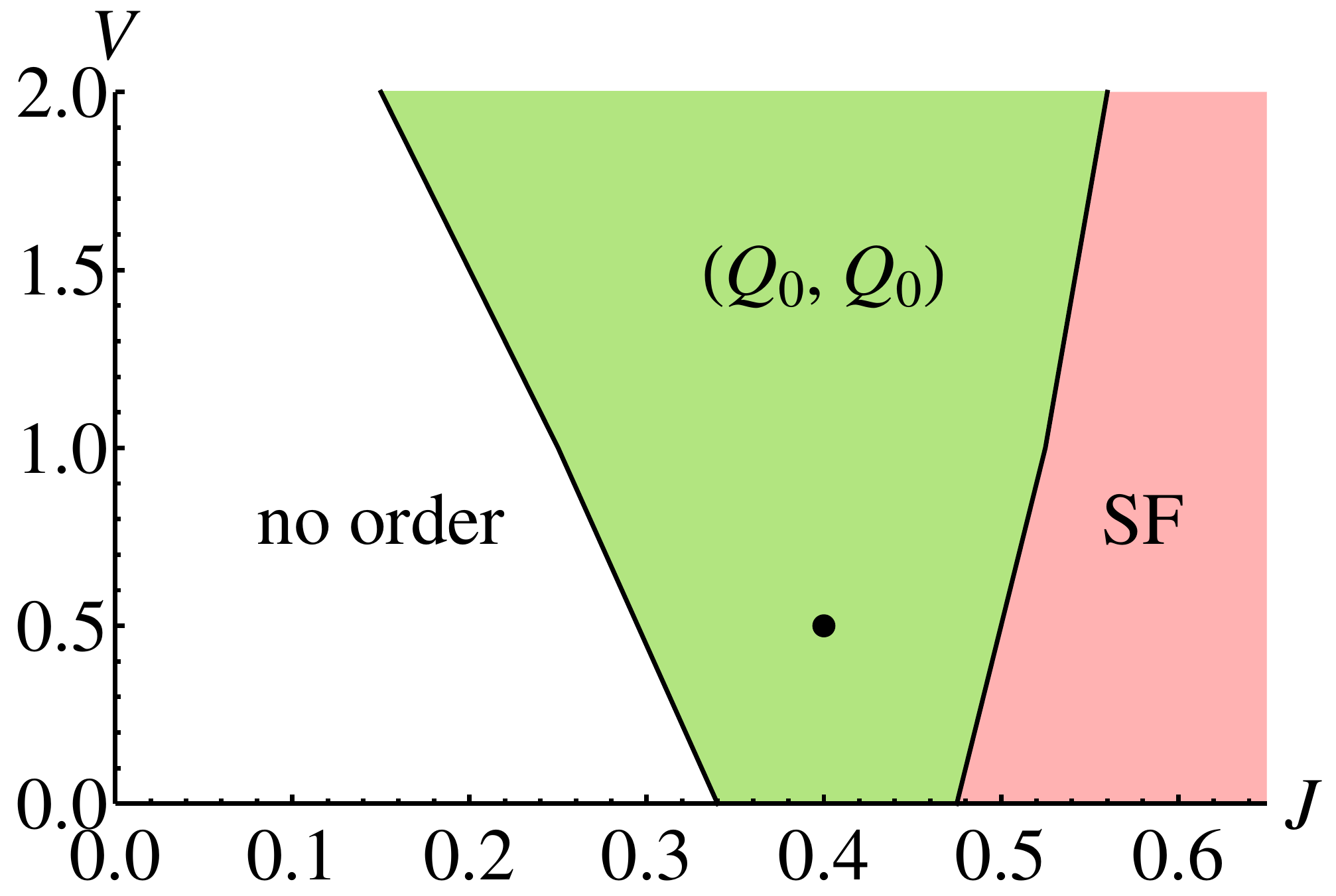}
\caption{\label{fig:variational} In the top row, the gain in variational energy
  (per site, times $100$) by allowing ordering at wavevector $\bQ$ is displayed. Each
  diagram is for a different Fermi surface, shown in the inset. The local
  minimum at $\bQ = (\pi, \pi)$ is the staggered flux state $P_5\neq 0$. The
  local minima at $\bQ = (Q_0, Q_0)$ and $\bQ = (Q_0, 0)$ are predominantly
  $d$-wave CDW ($P_3\neq 0$). The red dots denote wavevectors $\bQ$ that
  connect hot spots. In the bottom row, we show approximate diagrams of the ground
  state order, as a function of $J$ and $V$. The setup corresponding to the
  plot above is indicated with a black dot. Model parameters (values in the
  brackets correspond to plots from left to right): $x = 10\%$, $t_1
  = 1$, $t_2\in\{0.5,\,0.16,\,0.18\}$, $t_3\in\{0.6,\,0.9,\,1.6\}$, $J
  \in\{0.09,\, 0.235,\,0.4\}$, $V \in\{1.,\,1.5,\,0.5\}$. Variational
  parameters: $T_1 = 1.$, $T_2\in\{-0.1,\,-0.32,\,-0.5\}$, $T_3 = 0.128$. As
  described in the text the Fermi surface is determined by the $T_a$ parameters. 
}
\end{figure*}
We carry out the variational computation in a slightly unusual way. Ideally,
we would like to prescribe, and keep fixed, the Fermi surface of the system,
especially the presence and location of hot spots. This is the Fermi surface
seen in, say, photoemission experiments, and it is different from the one
obtained from the kinetic term in (\ref{ham}), because of interaction
effects. This ``true'' Fermi surface is not easily accessible, but the Fermi
surface of the optimal variational Hamiltonian is a close proxy. Therefore,
instead of minimizing with respect to the variational hoppings $T_a$, we set
them to values of our satisfaction, and tune the bare hoppings $t_a$ in such a
way as to make our choice a variational optimum \cite{jay}.  
We carry out this procedure while having all $P_i = 0$, then we look at what
CDW order can further improve the variational bound, while keeping $T_a$ fixed.
For each wavevector $\bQ$, we minimize
the variational energy, $\bra{\text{var}}H\ket{\text{var}}$, with respect to the parameters $P_{i}$, one at a
time. For a few choice wavevectors $\bQ$ we also looked at simultaneous
minimization with respect to $P_1$, $P_2$ and $P_3$, without finding
significant additional energy gains. This three parameter minimization is the
limit of the computational resources at our disposal. 

In order to have sufficient momentum resolution, we use a lattice of
$32\times32$ sites. Because the wavefunction is not a smooth function of the
variational parameters, we cannot use derivative information to carry out the
minimization, so we compute the energy over a grid of points in variational
space. More precisely, we store the expectation value of every operator making
up the Hamiltonian (\ref{ham}). This allows us to explore the parameter space
without having to recompute the energy for every setup. Our grid comprises a
total of about 30,000 points in variational space, each point requiring about
12 hr of CPU time on a 1.7GHz AMD Opteron.  

Fig.~\ref{fig:variational} illustrates the results of this variational
analysis. For a few representative choices of model parameters, we show  
the $\bvec Q$ values the system finds it energetically favorable to order. The
form factor corresponding to each energy minimum is not explicitly shown in
the plot. The minimum at $\bvec Q = (\pi, \pi)$ has $P_5\neq0$, i.e. it is the
SF state, whereas all other minima are predominantly $d$-wave CDW, i.e. have
$P_3 \neq 0$. The parameters are chosen close to the onset of the order. A
more comprehensive picture is given by the approximate phase diagrams of the
bottom row, which show the kind of order that yields the greatest gain in
energy over a range of parameters. The staggered flux state is dominant at
large $J$, but is suppressed by the Coulomb repulsion $V$ which prefers
CDWs. As can be seen in the upper row, in addition to this main order, other
subleading energy minima are usually present. In particular, even when the
global energy minimum is the staggered flux state, the CDWs are still present
as a local minimum. The CDWs become global minima at smaller $J$, 
and the appropriate Fermi surface configuration and moderate $J$
stabilize the wavevector $\bQ = (Q_0, 0)$ over $\bQ = (Q_0, Q_0)$. In our
framework, we cannot address the issue of competition between orders, as this
would require simultaneous minimization with respect to several more
parameters, but we expect that the $\bQ = (Q_0, 0)$ CDW can coexist with the
staggered flux state, since their mixing is prevented by time-reversal
symmetry.  

Our computations have shown that the observed charge order with $\bQ$ along
the copper-oxygen bonds appears over a regime of parameters in a variational
computation of a correlated metal with antiferromagnetic interactions. The
magnitude of $\bQ$  is close to that determined 
by the antiferromagnetic hot spots (Fig.~\ref{fig:bz}), and its form factor was robustly found to be predominantly $d$-wave
(defined as in Refs.~\cite{max,rolando}).

After our study was complete, we learnt of the related study of Ref.~\cite{troyer}
addressing symmetry breaking in the superconductor, rather than the metal, in a model
with only nearest-neighbor hopping.

We also note recent experimental reports \cite{comin2,dcdw} concluding that the 
charge order at $(Q_0,0)$ is predominantly $d$-wave, as discussed above.

\acknowledgments

We thank D. Chowdhury, A. Georges, and J. Sau for valuable discussions.
The research was supported by the U.S.\ National Science Foundation under
grant DMR-1103860, and by the Templeton Foundation. JB acknowledges financial
support from the DFG through grant number BA 4371/1-1. The simulation was done on
the MIT LNS Tier 2 cluster.


\begin{thebibliography}{}

\bibitem{tranquada} J.~Tranquada {\em et al.\/} Nature {\bf 375}, 561 (1995).

\bibitem{abbamonte} P.~Abbamonte {\em et al.\/} Nature Physics {\bf 1}, 155 (2005).

\bibitem{hoffman} J.~Hoffman {\em et al.\/} Science {\bf 295}, 466 (2002).

\bibitem{vershinin} M.~Vershinin {\em et al.\/} Science {\bf 303}, 1995 (2004).

\bibitem{julien} T.~Wu {\em et al.\/},
Nature {\bf 477}, 191 (2011).

\bibitem{julienvortex} T. Wu {\em et al.\/},
Nature Comm. {\bf 4}, 2113 (2013).

\bibitem{proust} D.~LeBoeuf, S.~Kr\"amer,	 W.~N.~Hardy, Ruixing Liang, D.~A.~Bonn, and C.~Proust,
Nature Physics {\bf 9}, 79 (2013).

\bibitem{doiron} N.~Doiron-Leyraud {\em et al.\/} Nature {\bf 447} 565 (2007).

\bibitem{sebastian} S.~Sebastian, N.~Harrison, and G.G.~Lonzarich, Rep. Prog. Phys. {\bf 75} 102501 (2012).

\bibitem{keimer} G.~Ghiringhelli {\em et al.\/}, 
Science {\bf 337}, 821 (2012).

\bibitem{hawthorn} A.~J.~Achkar {\em et al.\/},
Phys. Rev. Lett. {\bf 109}, 167001 (2012).

\bibitem{chang} J.~Chang {\em et al.\/},
Nature Phys. {\bf 8}, 871 (2012).

\bibitem{comin} R.~Comin {\em et al.\/} Science {\bf 343} 390 (2013).

\bibitem{ssrmp} S. Sachdev, Rev. Mod. Phys. {\bf 75}, 913 (2003).

\bibitem{kohsaka} Y.~Kohsaka {\em et al.}, {\it Science \/} {\bf 315}, 1380 (2007).

\bibitem{hawthorn2} A.~J.~Achkar {\em et al.\/},
Phys. Rev. Lett. {\bf 110}, 017001 (2013).

\bibitem{marston} I. Affleck and J. B. Marston, Phys. Rev. B {\bf 37}, 3774 (1988).

\bibitem{kotliar}  Z.~Wang, G.~Kotliar, and X.-F.~Wang, Phys. Rev. B {\bf 42}, 8690 (1990).

\bibitem{sudip} S.~Chakravarty, R.~B.~Laughlin, D.~K.~Morr, and C.~Nayak, Phys.
Rev. B {\bf 63}, 094503 (2001).

\bibitem{leewen} P. A. Lee, N. Nagaosa, and X.-G. Wen, Rev. Mod. Phys. {\bf 78}, 17 (2006).

\bibitem{laughlin} R.~B.~Laughlin, Phys. Rev. B {\bf 89}, 035134 (2014).

\bibitem{max} M.~A.~Metlitski and S.~Sachdev, Phys. Rev. B {\bf 82}, 075128 (2010).

\bibitem{rolando} S.~Sachdev and R.~La Placa, Phys. Rev. Lett. {\bf 111}, 027202 (2013).

\bibitem{metzner} T.~Holder and W.~Metzner, Phys. Rev. B {\bf 85}, 165130 (2012); 
C.~Husemann and W.~Metzner, Phys. Rev. B {\bf 86}, 085113 (2012).

\bibitem{yamase} M.~Bejas, A.~Greco, and H.~Yamase, Phys. 
Rev. B {\bf 86}, 224509 (2012).

\bibitem{pepin} K. B. Efetov, H. Meier, and C. P\'epin, 
Nature Physics {\bf 9}, 442 (2013).

\bibitem{kee} Hae-Young Kee, C.~M.~Puetter, and D.~Stroud, J. Phys.: Condens. Matter {\bf 25}, 202201 (2013).

\bibitem{bulut} S.~Bulut, W.~A.~Atkinson, and A.~P.~Kampf, Phys. Rev. B {\bf 88}, 155132 (2013).

\bibitem{dhlee} J.~C.~S\'eamus Davis and Dung-Hai Lee, Proc. Natl. Acad. Sci. {\bf 110}, 17623 (2013).

\bibitem{jay}  J. D. Sau and S. Sachdev, Phys. Rev. B {\bf 89},  075129 (2014).

\bibitem{pepin2} H. Meier, C. P\'epin, M. Einenkel, and K. B. Efetov, arXiv:1312.2010.

\bibitem{chubukov} Yuxuan Wang and A.~V.~Chubukov, arXiv:1401.0712.

\bibitem{gutz} M.~Gutzwiller, Phys. Rev. Lett. {\bf 10}, 159 (1963).

\bibitem{gros} C.~Gros, Ann. Phys. {\bf 189}, 53 (1989).

\bibitem{arun} A.~Paramekanti, M.~Randeria and N.~Trivedi, Phys. Rev. Lett. {\bf 87}, 217002 (2001)
and Phys. Rev. B {\bf 70}, 054504 (2004).

\bibitem{raczkowski} M.~Raczkowski, D.~Poilblanc, R.~Fr\'esard, and A.~M.~Ole\'s, Phys. Rev. B 75, 094505 (2007).

\bibitem{trivedi} S.~Pathak, V.~B.~Shenoy, M.~Randeria, and N.~Trivedi, Phys. Rev. Lett. {\bf 102}, 027002 (2009).

\bibitem{galatski} R.~Sensarma and V.~Galitski, Phys. Rev. B {\bf 84}, 060503(R) (2011).

\bibitem{stripes}  S.~A.~Kivelson, I.~P.~Bindloss, E.~Fradkin, V.~Oganesyan, J.~M.~Tranquada, A.~Kapitulnik, and C.~Howald,
Rev. Mod. Phys. {\bf 75}, 1201 (2003).

\bibitem{vojta4} M.~Vojta and O.~R\"osch, 
Phys. Rev. B {\bf 77}, 094504 (2008).

\bibitem{YK00}
H. Yamase and H. Kohno,
J. Phys. Soc. Jpn. {\bf 69}, 2151 (2000).

\bibitem{HM00}
C. J. Halboth and W. Metzner, Phys. Rev. Lett. {\bf 85}, 5162 (2000).

\bibitem{OKF01}
V. Oganesyan, S. A. Kivelson, and E. Fradkin,
Phys. Rev. B {\bf 64}, 195109 (2001).

\bibitem{varma} M. E. Simon and C. M. Varma, Phys. Rev. Lett.  {\bf 89}, 247003 (2002).

\bibitem{chetan} C. Nayak, Phys. Rev. B {\bf 62}, 4880 (2000).

\bibitem{troyer} P.~Corboz, T.~M.~Rice, and M.~Troyer, arXiv:1402.2859.

\bibitem{comin2} R.~Comin {\em et al.\/}, arXiv:1402.5415.

\bibitem{dcdw} K.~Fujita {\em et al.\/}, arXiv:1404.0362.









\end{thebibliography}
\end{document}